\documentclass[twocolumn,english,aps,prb,showpacs,byrevtex,amsmath,amssymb,superscriptaddress]{revtex4-1}
\usepackage{makecell}
\usepackage{dcolumn}
\usepackage{bm}
\usepackage{amsfonts}
\usepackage{latexsym}
\usepackage{color}
\usepackage{hyperref}
\usepackage[normalem]{ulem}
\usepackage{fancyhdr}
\usepackage{amsbsy}
\usepackage{amsmath}
\usepackage{subfigure}
\usepackage{epsfig}
\usepackage{amssymb}
\usepackage{pifont}
\usepackage{textcomp}
\usepackage{gensymb}
\usepackage{pifont}
\usepackage{enumerate}
\usepackage{color}
\usepackage{graphicx}
\usepackage[title]{appendix}

\usepackage{bm}
\usepackage{multirow}

\usepackage{tikz,xcolor,hyperref}
\definecolor{lime}{HTML}{A6CE39}
\DeclareRobustCommand{\orcidicon}{%
	\begin{tikzpicture}
	\draw[lime, fill=lime] (0,0)
	circle [radius=0.16]
	node[white] {{\fontfamily{qag}\selectfont \tiny ID}};
	\draw[white, fill=white] (-0.0625,0.095)
	circle [radius=0.007];
	\end{tikzpicture}
	\hspace{-2mm}
}

\foreach \x in {A, ..., Z}{%
	\expandafter\xdef\csname orcid\x\endcsname{\noexpand\href{https://orcid.org/\csname orcidauthor\x\endcsname}{\noexpand\orcidicon}}
}

\usepackage{hyperref}
\makeatletter
\newcommand{\printfnsymbol}[1]{%
  \textsuperscript{\@fnsymbol{#1}}%
}
\makeatother

\begin{document}

\begin{abstract}
We present a systematic study of the electronic and magnetic properties of two-dimensional ordered alloys, consisting of two representative hosts (MnPS$_3$ and NiPS$_3$) of transition metal phosphorus trichalcogenides doped with $3d$ elements.
For both hosts our DFT+U calculations are able to qualitatively
reproduce the ratios and signs of all experimentally observed magnetic couplings.
The relative strength of all antiferromagnetic exchange couplings, both in MnPS$_3$ as well as in NiPS$_3$, can successfully be explained using an effective direct exchange model:
they reveal that the third-neighbor exchange dominates in NiPS$_3$ due to the filling of the $t_{2g}$ subshell, whereas for MnPS$_3$ the first neighbor exchange is prevailed owing to the presence of the $t_{2g}$ magnetism. On the other hand, the nearest neighbor ferromagnetic coupling in NiPS$_3$ can only be explained using a more complex superexchange model and is (also) largely triggered by the absence of the $t_{2g}$ magnetism.
For the doped systems, the DFT+U calculations revealed that magnetic impurities
do not affect the magnetic ordering observed in the pure phases and thus in general in these systems ferromagnetism may not be easily induced by such a kind of elemental doping.
However, unlike for the hosts, the first and second (dopant-host) exchange couplings are of similar order of magnitude. This leads to frustration in case of antiferromagnetic coupling and may be one of the reasons of the observed lower magnetic ordering temperature of the doped systems.
\end{abstract}

\title{Limited ferromagnetic interactions in monolayers of MPS$_3$ (M=Mn, Ni)}


\author{Carmine Autieri
\orcidA}\thanks{These authors contributed equally.}
\affiliation{International Research Centre Magtop, Institute of Physics, Polish Academy of Sciences,
Aleja Lotnik\'ow 32/46, PL-02668 Warsaw, Poland}
\affiliation{Consiglio Nazionale delle Ricerche CNR-SPIN, UOS Salerno, I-84084 Fisciano (SA), Italy}

\author{Giuseppe Cuono\printfnsymbol{1}\orcidB}
\affiliation{International Research Centre Magtop, Institute of Physics, Polish Academy of Sciences,
Aleja Lotnik\'ow 32/46, PL-02668 Warsaw, Poland}

\author{Canio Noce\orcidF}
\affiliation{Dipartimento di Fisica ``E.R. Caianiello", Universit\`a degli Studi di Salerno, I-84084 Fisciano (SA), Italy}
\affiliation{Consiglio Nazionale delle Ricerche CNR-SPIN, UOS Salerno, I-84084 Fisciano (SA), Italy}

\author{Milosz Rybak}
\affiliation{Department of Semiconductor Materials Engineering, Faculty of Fundamental Problems of Technology, Wrocław University of Science and Technology, Wybrzeże Wyspiańskiego 27, PL-50370 Wrocław, Poland}

\author{Kamila M. Kotur}
\affiliation{Faculty of Physics, University of Warsaw, Pasteura 5, PL-02093 Warsaw, Poland}

\author{Cliò Efthimia Agrapidis\orcidC}
\affiliation{Faculty of Physics, University of Warsaw, Pasteura 5, PL-02093 Warsaw, Poland}

\author{Krzysztof Wohlfeld\orcidD}
\affiliation{Faculty of Physics, University of Warsaw, Pasteura 5, PL-02093 Warsaw, Poland}

\author{Magdalena Birowska\orcidE}
\affiliation{Faculty of Physics, University of Warsaw, Pasteura 5, PL-02093 Warsaw, Poland}
\maketitle

\section{Introduction}
Nonmagnetic van der Waals layered materials such as transition-metal dichalcogenides have been extensively studied over the last several years \cite{Manzeli2017}.
Just recently the intrinsic ferromagnetism in the true 2D limit has been reported \cite{Gibertini2019} initiating increasing excitement in the spintronics and 2D materials communities. In particular, the long range magnetic order has been observed for insulating CrI$_3$ \cite{Huang2017}, semiconducting Cr$_2$Ge$_2$Te$_6$ \cite{Gong2017} and metallic Fe$_3$GeTe$_2$ \cite{Deng2018} compounds. In addition, topological spin structures have been predicted for 2D materials\cite{Amoroso2020}.

Currently, the attention is on transition-metal phosphorus trichalcogenides (MPX$_3$, where M stands for transition atom and X=S, Se), which could be easily exfoliated down to monolayers \cite{Du2016, Lee2016} and are semiconducting materials with a wide range of band gaps \cite{arxiv2021}. 
The MPX$_3$ structures exhibit various antiferromagnetic arrangements within the magnetic ions, which are theoretically expected to be measurable  using different light polarization  \cite{PhysRevB.103.L121108}. 
Interestingly, the metal to insulator transition and superconductivity phase have been observed in this compound family  \cite{PhysRevB.100.214513,PhysRevLett.123.236401,PhysRevLett.120.136402}. In particular, the antiferromagnetic insulator phase in bulk FePS$_{3}$ can be melted and transformed into the superconducting phase under high pressure, providing similarity to the high-T$_c$ cuprate phase diagram \cite{Wang2018}. In addition,  theoretical predictions point to the existence of a large binding energy of excitons in  MnPS$_3$, whereas the experimental reports have observed excitons in few layers of NiPS$_3$ strongly related to magnetic order \cite{PhysRevLett.120.136402, Ho2021,Belvin2021}. Recent reports have demonstrated an all-optical control of the magnetic anisotropy in NiPS$_3$ by tuning the photon energy in resonance with an orbital transition between crystal-field split levels.\cite{Afanasieveabf3096}. The aforementioned demonstrates that this family of compounds is an ideal platform to study correlation effects in the true 2D limit.

In contrast to ferromagnetic materials, the antiferromagnets exhibit limited applications, mostly in the terahertz regime as ultra-fast components or specialized embedded memory-logic devices \cite{Nemec2018, Kampfrath2011, Kriegner2016}. Most of the current applications of magnetic crystals are based on the ferromagnetic (FM) semiconductors for which the band gap and FM order are crucial factors. The magnetic phase transitions for the MPX$_3$ materials can be accomplished by applying stress \cite{PhysRevB.94.184428}, changing the carrier concentration or applying voltage  \cite{Li2014}. In addition, the “M” atoms in MPX$_3$ crystals might be substituted with other transition metal atoms inducing the ferromagnetic order, as recently reported for a particular concentration of the non-magnetic dopants in the CrPSe$_3$ host resulting in half-metallic FM state \cite{adts.202000228}. Moreover, a series of mixed systems in a bulk form\cite{PhysRevB.104.174412,PhysRevB.54.14903,7c7fd3a4ca1,MASUBUCHI2008668,Goossens2000TheIO,Goossens_1998,V2000,Goossens2013,PhysRevMaterials.4.084401,HE200341,PhysRevMaterials.4.034411,PhysRevMaterials.5.064413}
have been experimentally realized, thus entering a new playground of magnetic phases. 

In addition, the magnetic and electronic properties of MPX$_3$ materials as well as critical N\'eel temperature (from T$_{N}$=78 K for MnPS$_3$\cite{Joy92,Siskins19,Long19} up to T$_{N}$=155 K for NiPS$_3$\cite{Joy92, Siskins19}) strongly depend  on the type of magnetic ion in the host. Thus, an elemental substitution could be an efficient way to tune the magnetic properties of atomically thin layers, by changing the lattice parameters and magnetic moments. These quantities result in manipulation of the exchange interactions that could be an effective way to engineer highly functional materials, similar to  magnetic heterostructures  \cite{Zhonge1603113}.  In particular, the main reason behind the idea that adding ions the partially filled \textit{d}-shells into the system might lead to the enhancement of the FM interactions is related to the so-called double exchange mechanism \cite{PhysRev.82.403}:
the hopping between two correlated ions with different valences (and
comparable Coulomb interactions) and relatively strong Hund's exchange
is allowed provided that the spins on the neighbouring ions is aligned
in a parallel fashion. Such a mechanism might for instance be at play
along the Cr-Mn bond: here the hopping from the manganese $e_g$ orbitals
($e_g^2$ configuration) to the chromium $e_g$ orbitals ($e_g^1$ configuration)
lowers the total energy of the system provided that the spins on Cr and
Mn are aligned ferromagnetic ally (for the antiparallel configuration
the strong Hund's exchange makes such a hopping energetically
unfavorable). Note that this double exchange mechanism is similar to
the one observed in the doped managanites which are ferromagnetic - albeit here it is a bit more complex, for it involves two different magnetic ions (nevertheless, the latter should not be that important,
for Cr and Mn have comparable values of the Coulomb interactions, cf.
Table II of \cite{PhysRevB.86.165105}).

Altogether, in this work we study the magnetic properties of the MPX$_3$ monolayers
and try to understand: (i) how ferromagnetic interactions can be stabilised in these
compounds, and (ii) whether one can easily modify these compounds by elemental doping so that
the ferromagnetic interactions can be strengthened. 
To this end, we examine two representative hosts MnPS$_3$ and NiPS$_3$ in the undoped case, as well as at a particular  doping concentration of magnetic ions. 
Using first principles calculations we give qualitative and quantitative explanations of the  origin of the exchange couplings strengths  up to the third nearest neighbors, and their respective ratios for MnPS$_3$ and NiPS$_3$ structures.
Considering the model hamiltonians with \textit{ab-initio} parametrization, we discuss the competition between the direct exchange  and the superexchange mechanisms  for the host structures. Next, we study various substantial sites of dopant atoms with mixed spins  and  mixed nearest-neighbor magnetic interactions. Here, again using first principles calculations, we examine in detail the mixed exchange coupling parameters between the metal host and dopant atoms.

\begin{table*}[ht]
\small
\caption{\label{tab:exchange}The exchange coupling strengths  $J_{i}^{\rm M}$ of MnPS$_3$ and NiPS$_3$ systems using model calculations as well as various
lattice parameters and on-site Coulomb repulsion $U$ in the DFT+U calculations. Positive (negative) $J_{i}^{\rm M}$ indicate AFM (FM) correlations, respectively. Note, that in Refs.~\cite{Wildes_1998,PhysRevB.98.134414} different convention of the exchange couplings $J_{i}^{\rm M}$ were used. The AFM-N and AFM-z indicate the magnetic ground state of the system. The exchange couplings in the model calculations are obtained within the direct exchange mechanism [for MnPS$_3$ using Eq. \ref{TOTEN_1} and Eq. \ref{TOTEN_3} whereas for NiPS$_3$ using Eq. \ref{TOTEN_2}]---except for $J_1^{\rm Ni}$ for NiPS$_3$ which contains also an important superexchange contribution and follows from Eq. \ref{eq7}, see text for further details. More details on the calculation of the magnetic exchanges within the DFT+U approach are present in Supporting Information. All values are given in meV.} 
 \def\arraystretch{1.3}
\begin{center}
\begin{tabular}{  l|  c|  c| c |c |c}
   \hline
\diaghead{\theadfont Diag ColumnmnHead II }%
{material MPS$_3$:}{  $J_i^\mathrm{M}$ [meV]}
 &\thead{$J_1^\mathrm{M}$}&  $J_2^\mathrm{M}$ & $J_3^\mathrm{M}$ & $|J_2^\mathrm{M}$/$J_1^\mathrm{M}|$& $|J_3^\mathrm{M}$/$J_1^\mathrm{M}|$\\
 \hline
MnPS$_3$ (model calculations, see caption above) &  19.5 & 0.35  &   7.76   & 0.02 & 0.40   \\
MnPS$_3$ (PBE, a=6.00{\AA} AFM-N)\cite{Olsen_2021} &  1.21& 0.18  &   0.54    &0.15 & 0.45   \\
MnPS$_3$ (U=5 eV, a=6.11{\AA}, AFM-N)  &1.04&0.05  &0.53 & 0.05& 0.51\\
MnPS$_3$ (U=5 eV, a=5.88{\AA}, AFM-N) \cite{PhysRevB.91.235425} & 1.58  & 0.08  &  0.46  &0.05 &0.29      \\
MnPS$_3$ (U=4 eV, a=6.00{\AA}, AFM-N) \cite{PhysRevB.94.184428} &  0.4& 0.03  &   0.15  &0.08&  0.38   \\
MnPS$_3$ (U=3 eV, a=6.00{\AA}, AFM-N) \cite{Olsen_2021} &  1.42& 0.08  &   0.52 &0.06 &0.37      \\
MnPS$_3$ (experiment, AFM-N) \cite{Wildes_1998} &  1.54& 0.14  &   0.36 & 0.09&  0.23    \\
\hline
NiPS$_3$ (model calculations, see caption above)  & -4.9& 0.06 & 14.8 & 0.01 & 3.0 \\
NiPS$_3$ (PBE, a=5.78{\AA}, AFM-z) \cite{PhysRevB.94.184428} & -11.3 &-0.12 &   36 & 0.01 & 3.2 \\
NiPS$_3$ (U=6 eV, a=5.84{\AA}, AFM-z)  &-3.34 &-0.19  &13.7 & 0.06 &  4.1  \\
NiPS$_3$ (U=4 eV, a=5.78{\AA}, AFM-z) \cite{PhysRevB.94.184428} & -4.11 & 1.95 &   17.4  & 0.47&  4.2 \\
NiPS$_3$ (U=3 eV, a=5.78{\AA}, AFM-z) \cite{Olsen_2021} & -2.6 & -0.32 &   14   &0.12 & 5.4 \\
NiPS$_3$ (experiment, AFM-z) \cite{PhysRevB.98.134414} & -3.8 & 0.2 &  13.8 & 0.05 & 3.6 \\
\hline
\end{tabular}
\end{center}
\end{table*}
\section{Results}
The results are presented as following: first,  we present the results for  the magnetic ground state of the hosts (pure MnPS$_3$, NiPS$_3$ systems). In particular, we consider the exchange couplings within the  DFT+U approach. Next, the antiferromagnetic (AFM) exchange mechanism is discussed within the  minimal many-body model, and the AFM exchange couplings strengths are evaluated numerically using the Wannier basis with \textit{ab-initio} parametrization. Next, the qualitative explanation of the ferromagnetic (FM) superexchange is presented.
Finally, we present comprehensive studies of electronic and magnetic properties of benchmark alloys with a fixed concetration of dopants. The elemental substitution is employed at various atomic sites of the honeycomb lattice.

\subsection{Undoped hosts MnPS$_3$ and NiPS$_3$}
\subsubsection*{Magnetic couplings within the DFT+U approach}
\begin{figure}
    \centering
    \includegraphics[width=0.5\textwidth]{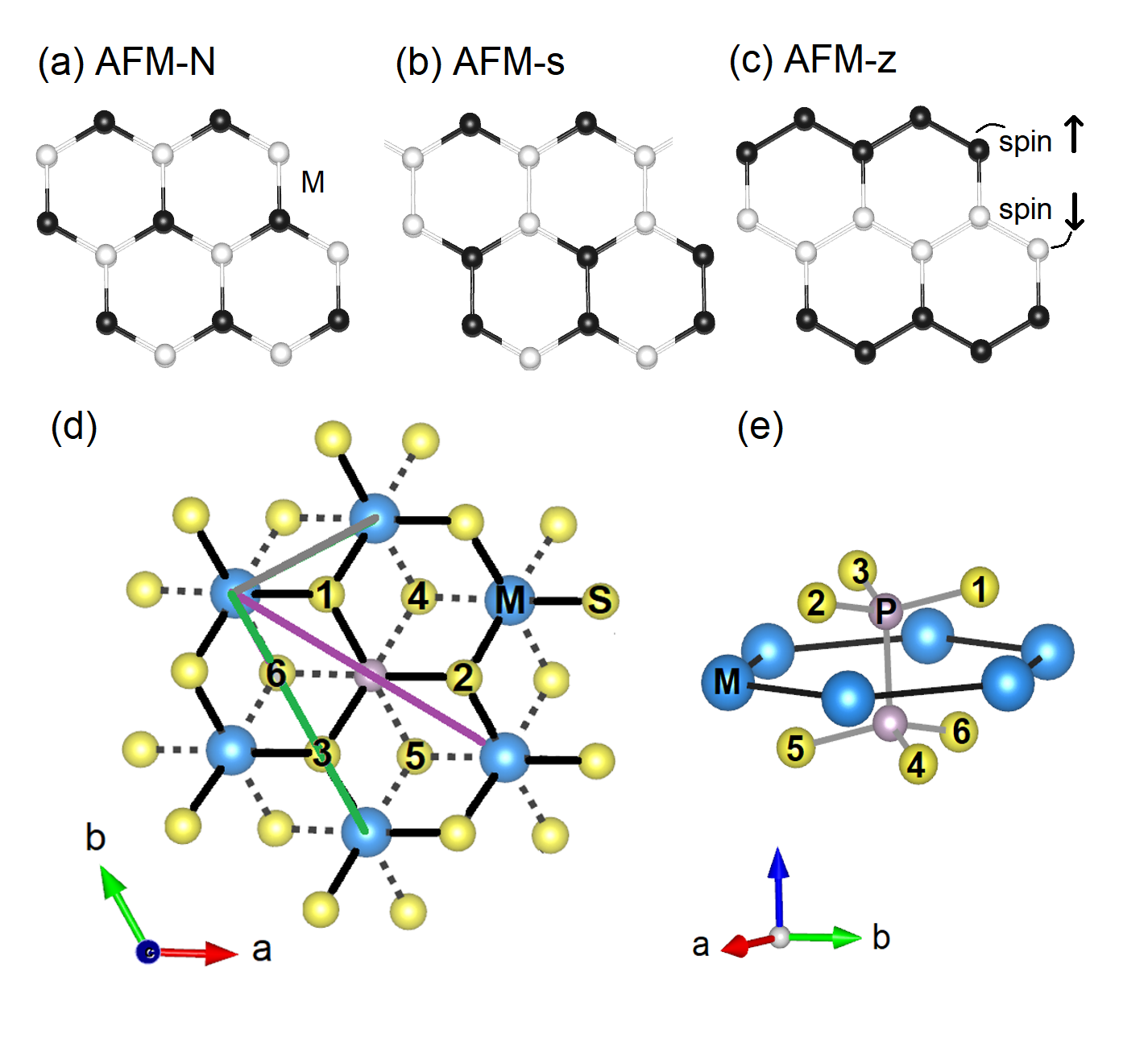}
    \caption{Spin arrangements of the metal  atoms in (a) AFM N\'eel (AFM-N), (b) AFM stripy (AFM-s) and (c) AFM zigzag (AFM-z) states. (d) Top and (e) side views of the crystal structure of MPX$_3$ system with the metal atoms denoted in blue, and surrounded by the sulphur atoms (yellow balls). The grey, green and violet lines indicate the NN, 2NN and 3NN distance between the metal atoms. The dotted and solid lines indicate that the the Sulphur atoms are below and above the metal layer, respectively.}
    \label{conf}
\end{figure}
The magnetic ground states of MnPS$_3$ and NiPS$_3$ exhibit AFM N\'eel (AFM-N) and AFM zigzag (AFM-z) type ordering (see Fig. \ref{conf}), respectively. The neutron diffraction data predicted that the Mn$^{2+}$ (3$d^5$, S=5/2) spins are slightly tilted (around 8$^\circ$) from the perpendicular direction of the honeycomb lattice \cite{PhysRevB.82.100408}, whereas the Ni$^{2+}$ (3$d^8$, S=1) spins are aligned within the honeycomb plane. In addition, due to the different filling of the \textit{3d} orbitals for various of metals ions (Fe, Mn, Ni etc.), also the size of the magnetic exchange coupling (J) changes. 
The existence of magnetic ordering at finite temperature in 2D limit requires the magnetic anisotropy  in accordance  to the Mermin-Wagner theorem. Currently only FePS$_3$, which posses a strong out-of plane easy axis, has been experimentally reported  to exhibit antiferromagnetic order in monolayer up to 118 K \cite{Lee2016}.
In order to explain the anisotropic order the  dipolar interactions between the magnetic moments or a single ion anisotropy resulting from a non zero spin-orbit coupling should be considered. The recent experimental reports, such as magnon band measurements  \cite{HICKS2019512}, support the claim that the dipolar interactions should be the leading term, whereas the electron spin resonance  \cite{CLEARY1986123} and  critical behaviour measurements \cite{Wildes2007Anis} indicate that the single ion anisotropy might come into play. In addition, experimental observations demonstrated that the antiferromagnetic ordering persists down to bilayer samples and is suppressed in the monolayer \cite{Kim2019}. Notably, the recent theoretical report \cite{KimNano21}, has questioned the Raman criterion used for the monolayer studies therein,  suggesting that NiPS$_3$ magnetic ordering could be presented in monolayer sample, as also indicated by strong two-magnon continuum existing in thin samples of NiPS3 \cite{Kim2019}. In the case of the MnPS$_3$ the magnetic order have been presented down to bilayer and was reported to be absent in the monolayer \cite{PhysRevLett.127.187201}. The suppression of the Neel temperature in thin samples can be associated with reducing the interplanar coupling, a in atomically thin samples \cite{Kim2019, Huang2017}. Aforementioned demonstrate that the magnetic ordering in monolayers of MPX$_3$ are still under a hot debate and many experiments are carrying out to verify the theoretical predictions. Similar effects of strong spin ﬂuctuation and absence of interlayer exchange coupling that weaken the long-range spin order in the 2D limit, have been reported in other layered magnets such as Cr$_2$Ge$_2$Te$_6$ and CrI$_3$ \cite{Huang2017}.

Here, we focus  on the the rationally invariant Heisenberg contribution, whereas the dipolar and single ion contributions are out of the scope of the present work and is widely discussed elsewhere \cite{KimNano21}. The latter has been estimated to 0.3 meV and 0.009 meV for NiPS$_3$ \cite{PhysRevB.98.134414} and MnPS$_3$ \cite{Wildes_1998}, respectively and discussed theoretically in \cite{Olsen_2021}.

The exchange interaction up to the third nearest neighbors ($J_{i}^\mathrm{M}$) between the metal  atoms of the hosts have been widely studied in a series of  previous works \cite{PhysRevB.91.235425, PhysRevB.94.184428, Olsen_2021, molecules26051410} (see Table \ref{tab:exchange}). Note that the prediction of the magnetic ground state within the DFT calculations does not depend on Hubbard interaction, whereas it is well known that the exchange coupling strength is sensitive to both U and the lattice parameters. We set the Hubbard U parameter to $U=5$ eV and $U=6$ eV for the 3$d$ orbitals of  Mn and Ni atoms, respectively.  These values are calculated from first principles using the Cococcioni approach \cite{Cococcioni05} (see Computation details). Our predicted  $J_i^\mathrm{M}$ values are in good agreement with  neutron diffraction experiments (see two last columns in Table \ref{tab:exchange}). The dominant exchange coupling is $J_3^\mathrm{Ni}$, which is much stronger than the $J_1^\mathrm{Ni}$. Note, that the experimental ratio of critical temperatures $T_N^{\mathrm{MnPS}_3}/T_N^{\mathrm{NiPS}_3}$ is reflected in the relation of dominant contributions $J_3^\mathrm{Ni}>J_1^\mathrm{Mn}$. In both cases, $J_2$ is much smaller than the other two exchange couplings. In particular, $J_2$ and $J_3$ couplings might be considered as superexchange interactions  involving the atoms in the path M-S1...S4-M for $J_2$ (see Fig. \ref{conf}d, \ref{conf}e), where the S atoms are located in different sublayers, whereas the $J_3$ interaction is mediated by S atoms located in the same sublayer through the bridge M-S1..S2-M. One could expect stronger hybridization of the S $p$ states and M 3$d$ states within the same sublayer of S atoms. In addition, the calculations reveal that $J_1$ is AFM and FM for MnPS$_3$ and NiPS$_3$, respectively. Note, that for both MnPS$_3$ and NiPS$_3$ the angle between the M-S-M atoms is close to 90$^\circ$ (83.4$^\circ$  for MnPS$_3$ and 85.4$^\circ$  for NiPS$_3$) for pointing to FM superexchange according to Goodenough-Kanamori-Anderson rules \cite{PhysRev.100.564,KANAMORI195987}. The direct M-M and indirect M-S-M (superexchange) mechanisms are of crucial importance to understand the differences between these two systems.

\subsubsection*{Effective direct exchange and antiferromagnetic couplings}
\begin{figure}[h!]
    \centering
    \includegraphics[width=0.5\textwidth]{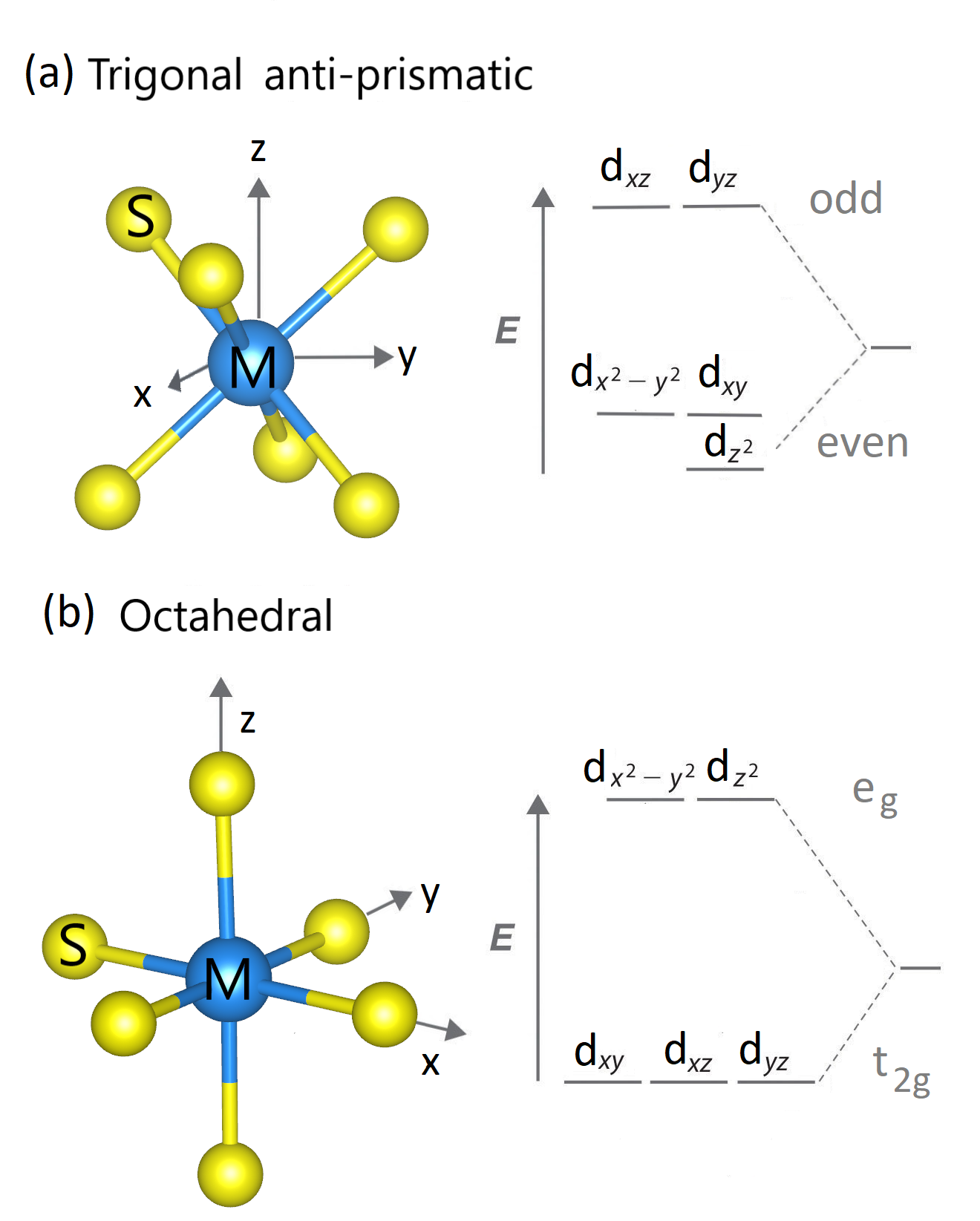}
    \caption{(a) Trigonal anti-prismatic  and (b) octahedral geometries. On the right side,  the \textit{d}-orbitals slitting of the metal atom in the corresponding crystal fields. M and S denote the metal atom and sulphur atom, respectively.}
    \label{symm}
\end{figure}
In order to understand the origin of the exchange couplings and their relative strengths we consider  model Hamiltonians for the direct and superexchange interactions. Note, that the direct exchange discussed throughout this work is a (second order) kinetic exchange process which involves only the hopping between  the transition metal ions, with the ligands not explicitly involved,  cf. \cite{Koch}, whereas a superexchange term is i.e. a fourth- (or higher-) order kinetic exchange
process which explicitly involves the hopping over the ligands,  cf. \cite{Koch}). In addition, we are pointing out the mechanisms which could impact the sign of $J_1$. 



To understand the origin of the magnetic couplings we first write down a minimal
many-body model which solely contains the valence electrons of the transition metal ion~\cite{PhysRevB.89.155109}---the multi-band Hubbard model. From that, using the second order perturbation theory that is valid in the Mott insulating limit, we derive the (effective) direct exchange processes. 
By construction all obtained spin couplings have to be antiferromagnetic. Therefore, while surprisingly successful, the following simple analysis will not be able to explain the onset of the ferromagnetic couplings in NiPS$_3$ (more on that in the end of the section). 

In the monolayers of MnPS$_3$ and NiPS$_3$, the metal atoms are surrounded by six 
sulphur atoms (MS$_6$ octahedron) and exhibit D$_{3d}$ point group symmetry in trigonal anti-prismatic environment of the ligands (sulphur atoms), see Fig. \ref{symm}(a). Hence, due to this trigonal crystal field effect, the $d$-manifold splits in two disentangled subsets of bands. 
The only coupling between these subsets is the spin-orbit coupling.\cite{Autieri2017}
The bands lower in energy are even ( $d_{x^2-y^2}$, $d_{xy}$ and $d_{z^2}$) with respect to the basal plane, while the  bands higher in energy are odd ($d_{xz}$ and $d_{yz}$) with respect to the basal plane. 
The Mn ion is $d^5$ and Mn $d$-bands split in half-filled even and half-filled odd bands, therefore the magnetic coupling acquires contributions both from the even and the odd orbitals.
Instead, since the Ni ion has a $d^8$ configuration, the even orbitals are fully occupied, therefore there is no magnetic contribution from these orbitals.
Altogether, the minimal model is the five-band Hubbard model with a simplified structure of the Coulomb interactions (no spin on-site spin exchange and pair-hopping terms, cf.~Ref.~\onlinecite{Anisimov1993}):
\begin{align}
\nonumber
H=& \sum_{im\sigma}\sum_{i^\prime m'\sigma'}\! 
   t^{i,i'}_{m,m'} u^{i,i'}_{\sigma,\sigma'} 
   c^{\dagger}_{im\sigma} c^{\phantom{\dagger}}_{i' m'\sigma'} \nonumber \\
&+U\sum_{ im }n_{im\uparrow }n_{im\downarrow} \nonumber\\
&+\frac{1}{2}\sum_{\substack{im( \neq m')\\ \sigma\sigma'}}(U-2J_H-J_H\delta_{\sigma,\sigma'}) n_{ im\sigma} n_{im'\sigma'}.
\label{H}
\end{align}
In this model $c_{im\sigma}^{\dagger}$ creates an electron with spin $\sigma\!=\uparrow,\downarrow$ in a Wannier orbital  $|m\rangle=|x^2-y^2\rangle$, $|xy\rangle$, $|xz\rangle$, $|yz\rangle$ or $|3z^2-r^2\rangle$  at site $i$, and $n_{im\sigma}=c_{im\sigma}^{\dagger}c^{\phantom{\dagger}}_{im\sigma}$. $\uparrow$ ($\downarrow$) indicates the spin up (down). 
The parameter $t^{i,i'}_{m,m'}$ is the hopping integral from orbital $m$ on site $i$ to  orbital $m'$ on site $i'$. The on-site terms $t_{m,m'}=\varepsilon_{m,m'}$ give the crystal-field splitting. $U$ and $J_H$ are the direct and (Hund) exchange  terms of the screened on-site Coulomb interaction. 

We perform second order perturbation theory in $t/U$ and for the commensurate electron filling of Mn and Ni ions. The direct exchange constant for the valence electrons occupying the Mn odd orbitals is:
\begin{equation}\label{TOTEN_1}
J_{i, i'}^{\mathrm{Mn},\text{odd}} \sim\frac{2(\mid t_{a,a}^{i,i'} \mid^2+\mid t_{b,b}^{i,i'} \mid^2) + \mid t_{a,b}^{i,i'} \mid^2 + \mid t_{b,a}^{i,i'} \mid^2}{U^\mathrm{Mn}+4J_H^\mathrm{Mn}},
\end{equation}
where $a$ and $b$ are the $d_{xz}$ and $d_{yz}$ orbitals, respectively, while $i,i'$ are the Mn-lattice sites.
In this formula we take in consideration both the cases with $i=1 \neq i'=2$ and $i=i'=1$.
$U^\mathrm{Mn}$ and $J_H^\mathrm{Mn}$ are the Coulomb repulsion and the Hund coupling in the case of the Mn atoms.
The Hund's rule interaction between odd and even electrons yields a magnetic coupling between these electrons, therefore the denominator depends on the occupancy of the even orbitals.
By symmetry the on-site energies are $\epsilon_{a}^{\mathrm{Mn}1}$=$\epsilon_{a}^{\mathrm{Mn}2}$=$\epsilon_{b}^{\mathrm{Mn}1}$=$\epsilon_{b}^{\mathrm{Mn}2}$.
\\

Similarly, we obtain the direct exchange constant for the valence electrons occupying the odd orbitals of the Ni atoms:
\begin{equation}\label{TOTEN_2}
J_{j,j'}^{\mathrm{Ni},\text{odd}}\sim\frac{2(\mid t_{a,a}^{j,j'}\mid^2+2\mid t_{b,b}^{j,j'}\mid^2)+\mid t_{a,b}^{j,j'}\mid^2+\mid t_{b,a}^{j,j'}\mid^2}{U^\mathrm{Ni}+J_H^\mathrm{Ni}},
\end{equation}
where $a$ and $b$ are the $d_{xz}$ and $d_{yz}$ orbitals, respectively, while $j,j'$ are the Ni-lattice sites.
In this formula we take in consideration both the cases with $j=1 \neq j'=2$ and $j=j'=1$.
$U^\mathrm{Ni}$ and $J_H^\mathrm{Ni}$ are the Coulomb repulsion and the Hund coupling in the case of the Ni atoms.
By symmetry the on-site energies are $\epsilon_{a}^\mathrm{Ni1}$=$\epsilon_{a}^\mathrm{Ni2}$=$\epsilon_{b}^\mathrm{Ni1}$=$\epsilon_{b}^\mathrm{Ni2}$.
\\

Finally, the direct exchange constant for the even Mn orbitals is the following:
\small
\begin{equation}\label{TOTEN_3}
\begin{split}
J_{i, i'}^{\mathrm{Mn},\text{even}}& \sim\frac{2(\mid t_{c,c}^{i,i'} \mid^2+\mid t_{d,d}^{i,i'} \mid^2 + \mid t_{e,e}^{i,i'} \mid^2) + \mid t_{d,e}^{i,i'}\mid^2 + \mid t_{e,d}^{i,i'}\mid^2}{U^\mathrm{Mn}+4J_H^\mathrm{Mn}}  
\\ &+ \frac{\mid t_{c,d}^{i,i'}\mid^2}{U^\mathrm{Mn}+4J_H^\mathrm{Mn}+\epsilon_{d}^{i'}-\epsilon_{c}^{i}}  + \frac{\mid t_{d,c}^{i,i'}\mid^2}{U^\mathrm{Mn}+4J_H^\mathrm{Mn}+\epsilon_{c}^{i'}-\epsilon_{d}^{i}} 
\\ & + \frac{\mid t_{c,e}^{i,i'}\mid^2}{U^\mathrm{Mn}+4J_H^\mathrm{Mn}+\epsilon_{e}^{i'}-\epsilon_{c}^{i}} + \frac{\mid t_{e,c}^{i,i'}\mid^2}{U^\mathrm{Mn}+4J_H^\mathrm{Mn}+\epsilon_{c}^{i'}-\epsilon_{e}^{i}},
\end{split}
\end{equation}
\normalsize
where $c$, $d$ and $e$ are the orbitals $d_{z^2}$, $d_{x^2-y^2}$, $d_{xy}$  respectively, while $i,i'$ are the metal lattice sites.
In this formula we take in consideration both the cases with $i=1 \neq i'=2$ and $i=i'=1$.
$U^\mathrm{Mn}$ and $J_H^\mathrm{Mn}$ are the Coulomb repulsion and the Hund coupling in the case of the Mn atoms.
By symmetry the on-site energies are $\epsilon_{c}^\mathrm{Mn1}$=$\epsilon_{c}^\mathrm{Mn2}$  and $\epsilon_{d}^\mathrm{Mn1}$=$\epsilon_{d}^\mathrm{Mn2}$=$\epsilon_{e}^\mathrm{Mn1}$=$\epsilon_{e}^\mathrm{Mn2}$.

The obtained this way (total) direct exchange, $J=J^\text{odd}+J^\text{even}$,
is positive for both Mn and Ni ion and for any distance $j-j'$. Hence, as already mentioned, it is always antiferromagnetic by construction.

\subsubsection*{Numerical evaluation of the antiferromagnetic couplings}

The Wannier basis provides us with {\em ab-initio} values of the hopping integrals and crystal-field splittings.
We calculate the hopping parameters and the on-site energies using the interpolated band structure of the Wannier functions of the $d$-subsector.
The on-site energies for Mn and Ni are: $\epsilon_{a}^\mathrm{Mn1}$=$-$1197.7 meV,  $\epsilon_{c}^\mathrm{Mn1}$=$-$2082.6 meV, $\epsilon_{d}^\mathrm{Mn1}$=$-$2179.3 meV and $\epsilon_{a}^\mathrm{Ni1}$=$-$1768.5 meV.
We have three first nearest neighbours (1NN), six second nearest neighbours (2NN) and three third nearest neigbours (3NN).
In the case of  odd orbitals, the 3NN couplings are greater than the 1NN couplings, even by an order of magnitude; therefore it is very important to consider these hopping amplitudes in the calculations.
On the other hand, for  even orbitals, the 1NN couplings are greater than the 3NN couplings.
The 2NN are always smaller with respect to the 1NN couplings and the 3NN couplings. 
The 3NN couplings are important also by account of the Mermin-Wagner theorem, that prohibits long-range magnetism when only in-plane first and second nearest neighbors are concerned \cite{PhysRevLett.17.1133,Noce06}.
In the case of  even orbitals we neglect the difference between the  on site energies assuming the following approximation $\epsilon_{c}^\mathrm{Mn}-\epsilon_{d}^\mathrm{Mn}$, $\epsilon_{c}^\mathrm{Mn}-\epsilon_{e}^\mathrm{Mn}\ll(U^\mathrm{Mn}+4J_H^\mathrm{Mn})$. \\

Now, we will numerically evaluate  the second- and third-neighbour direct exchange as a function of the first-neighbour direct exchange. 
For the odd subsector of the MnPS$_3$, we obtain that
$J_2^\mathrm{Mn,odd}=0.037 J_1^\mathrm{Mn,odd}$ and  $J_3^\mathrm{Mn,odd}=2.026 J_1^\mathrm{Mn,odd}$.
For the even subsector of the MnPS$_3$, we obtain that
$J_2^\mathrm{Mn,even}=0.015 J_1^\mathrm{Mn,even}$ and  $J_3^\mathrm{Mn,even}=0.016 J_1^\mathrm{Mn,even}$.
Considering that $J^\mathrm{Ni,even}=0$ due to fully occupied even orbitals, for the odd subsector of  NiPS$_3$, we obtain that
$J_2^\mathrm{Ni,odd}=0.047 J_1^\mathrm{Ni,odd}$ and  $J_3^\mathrm{Ni,odd}=11.09 J_1^\mathrm{Ni,odd}$.
If we consider the total direct exchange value as the sum of the odd and even sector, we have 
$J_1^\mathrm{Mn}=(152.4 \text{eV} \cdot \text{meV})/(U^\mathrm{Mn}+4J_H^\mathrm{Mn})$ for the first neighbour coupling in MnPS$_3$. For the second- and third-neighbours, we obtain
$J_2^\mathrm{Mn}=0.018 J_1^\mathrm{Mn}$ and
$J_3^\mathrm{Mn}=0.397 J_1^\mathrm{Mn}$, therefore, for  MnPS$_3$, the dominant direct exchange comes from the first-neighbour coupling. Using the Coulomb repulsion of 5 eV for Mn and an Hund coupling of 0.7 eV, we obtain the numerical values equal to J$_1^{Mn}$=19.5 meV, J$_2^{Mn}$=0.35 meV and J$_3^{Mn}$=7.76 meV, (see Table \ref{tab:exchange}.)
 
When we numerically evaluate  the direct exchange of NiPS$_3$, we obtain
$J_1^\mathrm{Ni}=(8.957 \text{eV} \cdot \text{meV})/(U^\mathrm{Ni}+J_H^\mathrm{Ni})$ for the first-neighbour coupling, 
$J_2^\mathrm{Ni}=0.047 J_1^\mathrm{Ni}$ and
$J_3^\mathrm{Ni}=11.09 J_1^\mathrm{Ni}$ for the second- and third-neighbours, respectively.
Remarkably, the 3NN magnetic direct exchange is larger than the 1NN in the odd case. Using the Coulomb repulsion of 6 eV for Ni and an Hund coupling of 0.7 eV, we obtain the numerical values equal to J$_1^{Ni}$=1.3 meV, J$_2^{Ni}$=0.06 meV and J$_3^{Ni}$=14.8 meV, where last two are reported in Table \ref{tab:exchange}.
\\

Altogether, we obtain that the simple direct exchange scheme gives
$J_3^\mathrm{Ni}\gg J_1^\mathrm{Ni}\gg J_2^\mathrm{Ni}$ and
$J_1^\mathrm{Mn}>J_3^\mathrm{Mn}\gg J_2^\mathrm{Mn}$.
In the Ni case, the leading term is $J_3^\mathrm{Ni}$ while in the Mn case the leading term is $J_1^\mathrm{Mn}$.
The reason for this different behaviour comes from the different filling that produces $J^\mathrm{Ni,even}=0$.
The calculated direct exchange couplings are qualitatively in agreement with the magnetic couplings obtained experimentally or using DFT---except that the $J_1^\mathrm{Ni}$ and $J_2^\mathrm{Ni}$ are ferromagnetic due to more complex magnetic exchange not taken into account by the (simple) direct exchange scheme (see below). 
Even though the latter coupling are relatively small, note that considering in more detail such discrepancy is relevant for an accurate description of the magnetic coupling and, hence, of the magnetic critical temperature.
The different leading exchange terms in Mn and Ni compounds open the way to manipulating the magnetism by tuning the concentration of Mn and Ni compounds, or by adding new magnetic materials as dopants.


\begin{figure*}[t!]
\centering
\includegraphics[width=0.98\textwidth]{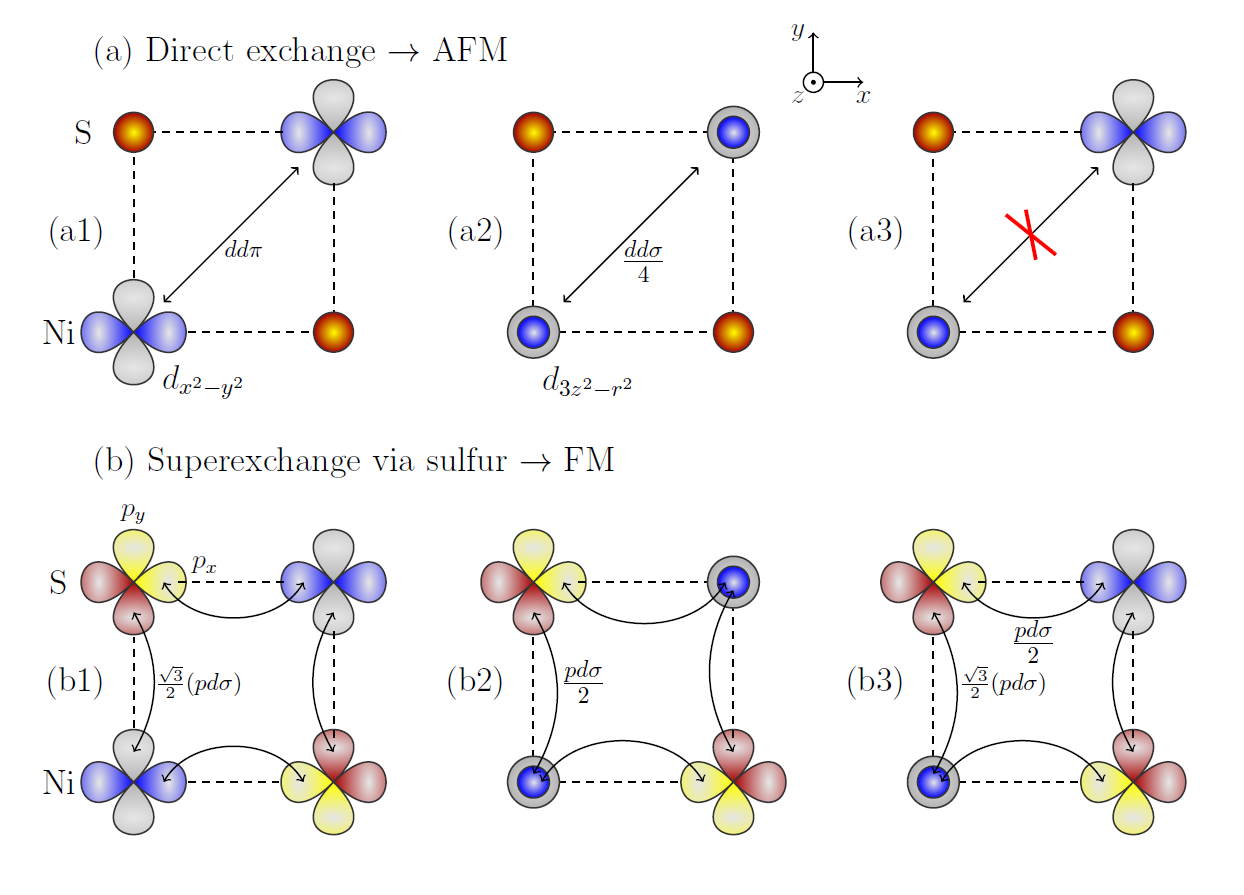}
\caption{\label{pdexchange} Schematic comparison between the AFM direct exchange (top panels) and the FM superexchange (bottom panels) processes in NiPS$_3$. Finite direct AFM exchange processes due to: (a1) nonzero hopping elements $\propto dd\pi$ between the nearest neighbor $d_{x^2-y^2}$ orbitals on nickel and (a2)  nonzero hopping elements $\propto dd\sigma $ between the nearest neighbor $d_{3z^2-r^2}$ orbitals on nickel. Lack of a direct AFM exchange processes due to vanishing hopping elements between the nearest neighbor $d_{x^2-y^2}$ and $d_{3z^2-r^2}$ on nickel. Finite FM superexchange processes due to: (b1) nonzero hopping elements $\propto pd\sigma$ between the nearest neighbor $d_{x^2-y^2}$ orbitals on nickel and the nearest neighbor $p_x / p_y$ orbital on sulfur, (b2) nonzero hopping elements $\propto pd\sigma$ between the $d_{3z^2-r^2}$ orbitals on nickel and the nearest neighbor $p_x / p_y$ orbital on sulfur, (b3) nonzero hopping elements $\propto pd\sigma$ between the $d_{x^2-y^2}$ orbital on nickel and the nearest neighbor $p_x / p_y$ orbital on sulfur as well as the $d_{3z^2-r^2}$ orbital on nickel and the nearest neighbor $p_x / p_y$ orbital on sulfur. See text for more details.}
\end{figure*}

\subsubsection*{Superexchange and a ferromagnetic coupling in NiPS$_3$}

So far we considered an effective direct exchange model, which solely contained the exchange processes due to the electrons hopping between the transition metal (Mn or Ni) ions. Note that such a model should be considered an effective one, for in reality the hopping between the neighboring Mn or Ni ions is predominantly mediated by the sulfur ions. Hence, a natural extension of the direct exchange model should explicitly contain the exchange processes on the sulfur ions---in fact, it is such a nearest neighbor superexchange model~\cite{Ushakov2013} that is studied below to explain the onset of the nearest neighbor ferromagnetic exchange in NiPS$_3$.
Note that for consistency we comment in the end of this subsubsection why the more complex superexchange model is not needed 
to understand the other, the first- and third-neighbor~\footnote{The second neighbor coupling is very small and we leave it out from the analysis.}, magnetic couplings in MPS$_3$---i.e. that the effective direct exchange model is enough.

We introduce the nearest neigbbor superexchange model~\cite{Ushakov2013} for NiPS$_3$ by considering two Ni ions connected via two sulfur atoms over two 90 degree bonds, see Fig.~\ref{pdexchange}. 
Note that, since Ni$^{2+}$ is in a $d^8$ configuration and sulfur ions are fully occupied, it is easier to consider the hole language in below.
Moreover, in what follows we neglect the small trigonal distortions and we assume an octahedral crystal field with the division in t$_{2g}$ and e$_g$ orbitals [with the coordinate system defined in such a way that the $xy$ plane coincides with the plane formed by the nearest neighbor transition metal ion and sulfur, see Fig.~\ref{symm}(b) and Fig.~\ref{pdexchange}].
Hence we begin by considering a fully atomic limit without hopping (0th order of perturbation theory in the small kinetic energy) in which there are two holes localised in two distinct Ni $e_g$ orbitals ($d_{x^2-y^2}, d_{3z^2-r^2}$; in what follows assumed to be energetically degenerate, see also above) and two (higher-lying by the charge transfer energy $\Delta$) empty $p$ orbitals ($p_x, p_y$) on sulfur. As before, due to the strong Hund's rule $J_H$ the two Ni$^{2+}$ holes form a high-spin $S=1$ state. Now let us perform a perturbation theory in the kinetic energy (over Coulomb repulsion $U$ and charge transfer energy $\Delta$) and consider the possible exchange processes---which are of two kinds:

First, there are the direct exchange processes between
the nickel ions, see Fig.~\ref{pdexchange}(a). By definition these concern virtual occupancies of one of the nickel ions by three holes [with a relative energy cost of $ U+J_H$ according to the simplified structure of the Coulomb interactions, see Eq.~(\ref{H})] and are possible once the hole can directly hop back and forth between the nickel orbitals under consideration. According to the Slater-Koster scheme~\cite{Slater1954}, which is qualitatively confirmed by our DFT calculations, the latter is allowed between the pair of $d_{x^2-y^2}$ orbital ($dd\pi$ hopping element) and between the pair of $d_{3z^2-r^2}$ orbitals ($dd\sigma/4$ hopping element; the small $dd \delta$ element can be neglected), cf. Fig.~\ref{pdexchange}(a1-a2); note that the hopping between the $d_{x^2-y^2}$
and the $d_{3z^2-r^2}$ orbital vanishes in this geometry, cf. Fig.~\ref{pdexchange}(a3). Altogether we obtain the nearest neighbor direct exchange contribution
\begin{align}\label{direct}
J^{\text{Ni}, \text{direct}}_1 = 4 \left[\frac{(dd\pi)^2}{U+J_H} + \frac{(dd\sigma/4)^2}{U+J_H}\right] \approx 4\frac{(dd \pi)^2}{U+J_H},
\end{align}
where we assumed a typical relation between the Slater-Koster hopping integrals $dd\sigma \approx 2 dd \pi $.
It is important to state here that the above direct exchange process is different than the {\it effective} direct exchanges process defined in the previous subsections---for the latter ones may include all indirect hoppings (e.g. via sulfur) between the two nearest neighbor nickel ions.

Second, there are the superexchange processes between the nickel ions, see Fig.~\ref{pdexchange}(b). By definition these concern virtual occupancies of one of the sulfur ions by two holes (with an energy cost associated with the charge transfer $2\Delta$ for antiparallel spins or $2\Delta - J_H$ for parallel spins~\footnote{Note that the finite Coulomb repulsion cost on sulfur just renormalizes the (anyway quite hard to estimate) charge transfer energy.}) and are possible once the two holes hop back and forth between the sulfur and nickel orbitals under consideration. Again, using the Slater-Koster scheme~\cite{Slater1954},
and considering the three distinct possible hopping processes shown in Figs.~\ref{pdexchange}(b1-b3), we obtain the nearest neighbor superexchange contribution
\begin{align}
\label{SE}
J^{\text{Ni}, \text{SE}}_1 &= \frac12 \left(\frac{4}{2\Delta} - \frac{4}{2\Delta - J_H} \right)\nonumber \\
&\times \Bigg[
\left( \frac{\sqrt{3} pd \sigma}{2} \right)^2 \left( \frac{\sqrt{3} pd \sigma}{2} \right)^2 \frac{2}{\Delta^2} \nonumber \\
&+
\left( \frac{ pd \sigma}{2} \right)^2 \left( \frac{ pd \sigma}{2} \right)^2
\frac{2}{\Delta^2}
 \nonumber \\
&+ \left( \frac{ \sqrt{3} pd \sigma}{2} \right)^2  \left( \frac{ pd \sigma}{2} \right)^2
\frac{4}{\Delta^2} \Bigg]^2 \nonumber \\
 &= -2 J_H  \frac{(pd\sigma)^4 }{  (2 \Delta - J_H) \Delta^3},
\end{align}
which is negative (ferromagnetic) due to the lowering of the energy for a ferromagnetic 
(virtual) occupancy of the sulfur by two holes in the superexchange process. 
Note that for a given Ni-Ni bond there are always two superexchange processes: one over the top-left and one over the bottom-right sulfur (hence the factor of two within the square brackets in the above formula). Moreover, the process depicted by Fig.~\ref{pdexchange}(b3) has to be multiplied by a factor of two, for one can interchange the position of the $d_{x^2-y^2}$ 
and the $d_{3z^2-r^2}$ orbital and this way double the amplitude of this process.
Finally note that, due to the fact that we have spins $S=1$ on nickel, overall the above superexchange process is reduced by a factor $1/2$ w.r.t. to an analogous one for the $S=1/2$ on copper (the superexchange processes have to be projected onto the high spin $S=1$ states
on both nickel ions---hence a factor of $(1/\sqrt{2})^2$ reduction).

Let us now comment why the above ferromagnetic superexchange mechanism is not important for the nearest neighbor exchange in MnPS$_3$. The reason for this is that in the case of manganese ions we are a bit closer to the situation discussed in Ref.~\onlinecite{Ushakov2013}, which for instance shows the antiferromagnetic exchange coupling in the case of half-filled $t_{2g}$ subshell of Cr$^{3+}$ in LiCrS$_2$. More precisely, the situation for the Mn$^{2+}$ ions is as follows. On one hand, the antiferromagnetic direct exchange is much stronger for Mn$^{2+}$, since one of the $t_{2g}$ electrons (the $d_{xy}$) can hop over the $dd\sigma$ bond. On the other hand, the superexchange also contains an additional strong~\cite{Ushakov2013} antiferromagnetic contribution, due to the superexchange processes over one $p_z$ sulfur orbital---which strongly hybridizes with two nearest $t_{2g}$ orbitals. Altogether, as confirmed by the effective direct exchange studies in the previous subsection, these two mechanisms originating in the $t_{2g}$ exchange processes easily overcome the (above-described) ferromagnetic processes for the $e_g$ orbitals. 

Finally, we mention that the (surprisingly) strong third-neighbor coupling in MPS$_3$
can easily be explained by the superexchange model. First, for the $t_{2g}$ sector
the strong antiferromagnetic third-neighbor coupling is already discussed in Ref.~\onlinecite{Ushakov2013}---see Fig.~7 of that reference. Second, one can easily imagine that 
a similarly strong antiferromagnetic coupling can also be realised for the $e_g$ electrons:
in order to understand it, one just needs to replace the two third neighbor $d_{xy}$ orbitals by the $d_{x^2-y^2}$ orbitals and rotate the sulfur $p$ orbitals by 90 degrees 
in the process shown in Fig.~7 of Ref.~\onlinecite{Ushakov2013}. In fact, the latter
one should have a very high amplitude and hence a really strong third-neighbor exchange
in NiPS$_3$.

\subsubsection*{Numerical evaluation of the ferromagnetic coupling in NiPS$_3$}

Having derived the direct exchange and superexchange processes between the nearest neighbor nickel ions, we are now ready to estimate the contributions of these both (competing) spin interaction terms. To this end, we assume that the Coulomb repulsion $U= 6$ eV (the chosen value of our DFT+U approach, see Table I) and take a typical (for $3d$ transition metal compounds) value of the Hund's exchange $J_H = 0.7$ eV. Next, based on the DFT calculations we estimate that:
(i) the charge transfer energy $\Delta = 3 $ eV, (ii) the hopping $pd \sigma  = 0.9 $ eV,
and (iii) the hopping $dd \pi = 0.05$ eV. From that, and using Eq.~(\ref{direct}) and Eq.~(\ref{SE})
we can easily calculate the antiferromagnetic and ferromagnetic contributions to the spin exchange:
\begin{align}\label{eq7}
    J^{\text{Ni}, \text{direct}}_1 \approx 1.5~{\rm meV}, \quad J^{\text{Ni}, \text{SE}}_1 \approx -6.4~{\rm meV}.
\end{align}
As the second exchange is larger (by absolute value), we conclude that it is the relatively strong superexchange along the 90 degree nickel-sulfur-nickel bonds which triggers the ferromagnetic exchange along the nearest neighbor nickel ions.

At this point we would like to comment on the fact that for the edge sharing copper chains,
typically one entirely neglects the direct exchange~cf.~Ref.~\onlinecite{Johnston2016}. The reason for that is that, in the copper chains, the direct copper hopping  $dd \sigma = 0.08$ eV~\cite{Rosner1999}, whereas usually one assumes that for the cuprates $p d \sigma \approx 1.5 $ eV. With such hopping parameters (and assuming the typical cuprate values of $U=8$ eV, $\Delta = 3$ eV, and oxygen $J_H= 0.7$ eV), one can immediately see that for the copper chains with 90 degree geometry $ J^{\text{Cu}, \text{direct}} \approx 3 {\rm meV}  \ll |J^{\text{Cu}, \text{SE}}| \approx 50 {\rm meV}$---which justifies why the direct copper exchange is typically neglected in such studies. (In reality the ferromagnetic exchange is actually really small in the such copper chains, due to the angle along the copper-oxygen-copper bond not being strictly equal to 90 degrees). 
A strong suppression o the direct exchange is also observed in CrI$_3$ \cite{doi:10.1063/5.0039979}, which is an experimentally confirmed  2D ferromagnet.  CrI$_3$ is isostructurally identical to MPS$_3$ compounds when P atoms are removed
and has partially-filled $t_{2g}$ shell, just as MnPS$_3$. 
This might suggest that CrI$_3$ should predominantly have antiferromagnetic interactions, just as MnPS$_3$.
However, this is not the case, since the relative strength of the direct and superexchange mechanisms is different in CrI$_3$ and MnPS$_3$. In CrI$_3$, the Cr-I-Cr bond angle is very close to 90 degrees leading to a strong FM superexchange while the Cr-Cr distance is relatively large (3.95 $\AA$) giving rise to a very weak AFM direct exchange. Interestingly, due to the larger distances between the magnetic atoms in CrI$_3$, the third-neighbor exchange is also less relevant.

\subsection{Doped systems (M$_{3/4}$,X$_{1/4}$)PS$_3$ }
\begin{figure*}[th]
\centering
\includegraphics[width=1.0\textwidth]{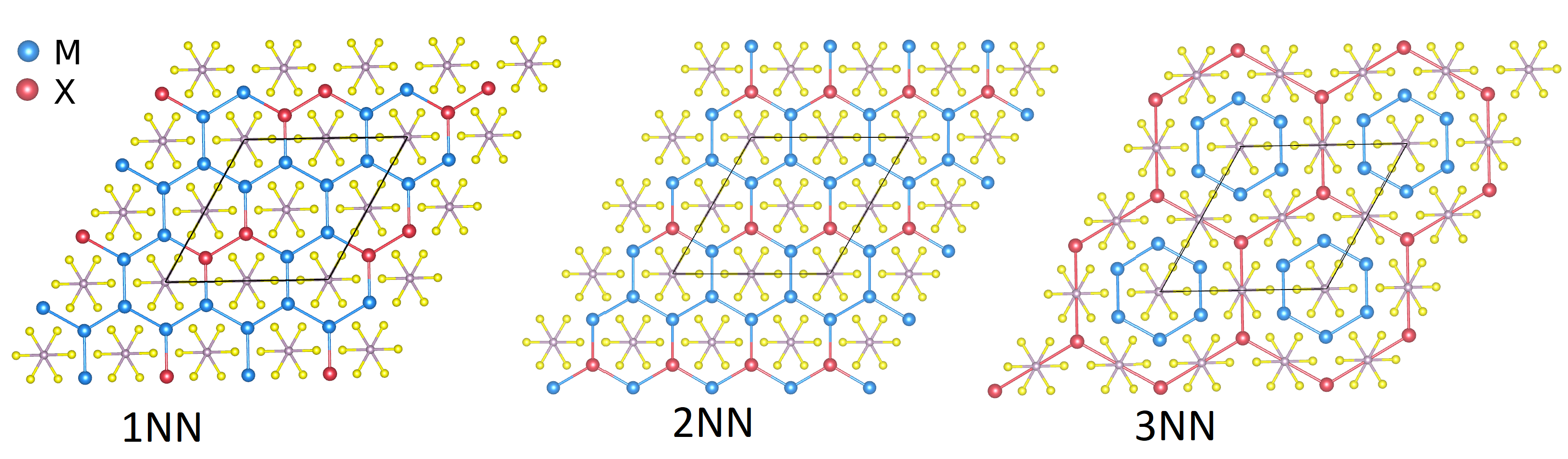}
\caption{\label{supercell}  Crystal structure of $2\times2$ supercell consisting of 4-fold primitive unit cells.} The M, X,  are the transition metals of the host and dopant atoms, respectively.  The 1NN, 2NN,and 3NN denote the nearest neighbours, second neighbours and third neighbours of the dopants, respectively.
\end{figure*}

Our strategy is to try to induce long-range ferromagnetic order via chemical doping keeping the two-dimensional structure of the mother compounds intact.  In principle, the chemical doping could influence the magnetic exchanges, especially the long-range ones.
To this end, we consider doping of the host systems with distinct possible $3d$-elements. In particular, at low doping concentration, the main exchange couplings are the magnetic exchanges between the host-magnetic atoms ($J_i^\mathrm{M}$ for $i=1,2,3$) and between the host-impurity  atoms ($J_i^\mathrm{MX}$ for $i=1,2,3$), respectively. While for most of the $3d$-doping the Mn-Mn and the Ni-Ni magnetic couplings remain antiferromagnetic, in the case of  Cr-doping the antiferromagnetic exchanges between the magnetic atoms of the host turn into ferromagnetic ones \cite{Son2021}.
Thus, we performed DFT+U calculations for 25$\%$ of the  concentration  of the magnetic dopants (X=Cr, Mn, Ni) in the hosts of NiPS$_3$ and MnPS$_3$. We have employed various structural arrangements of the atoms (see Fig. \ref{supercell}), and collinear spin configurations such as antiferromagnetic Neel (AFM-N), zig-zag (AFM-z), stripy (AFM-s) and ferromagnetic case (FM). For each of the configuration the atomic positions have been fully optimized, keeping the lattice constants equal to the pure optimized monolayer structures (see Appendix A).

 
 
\begin{figure}
\centering
\includegraphics[width=0.5\textwidth]{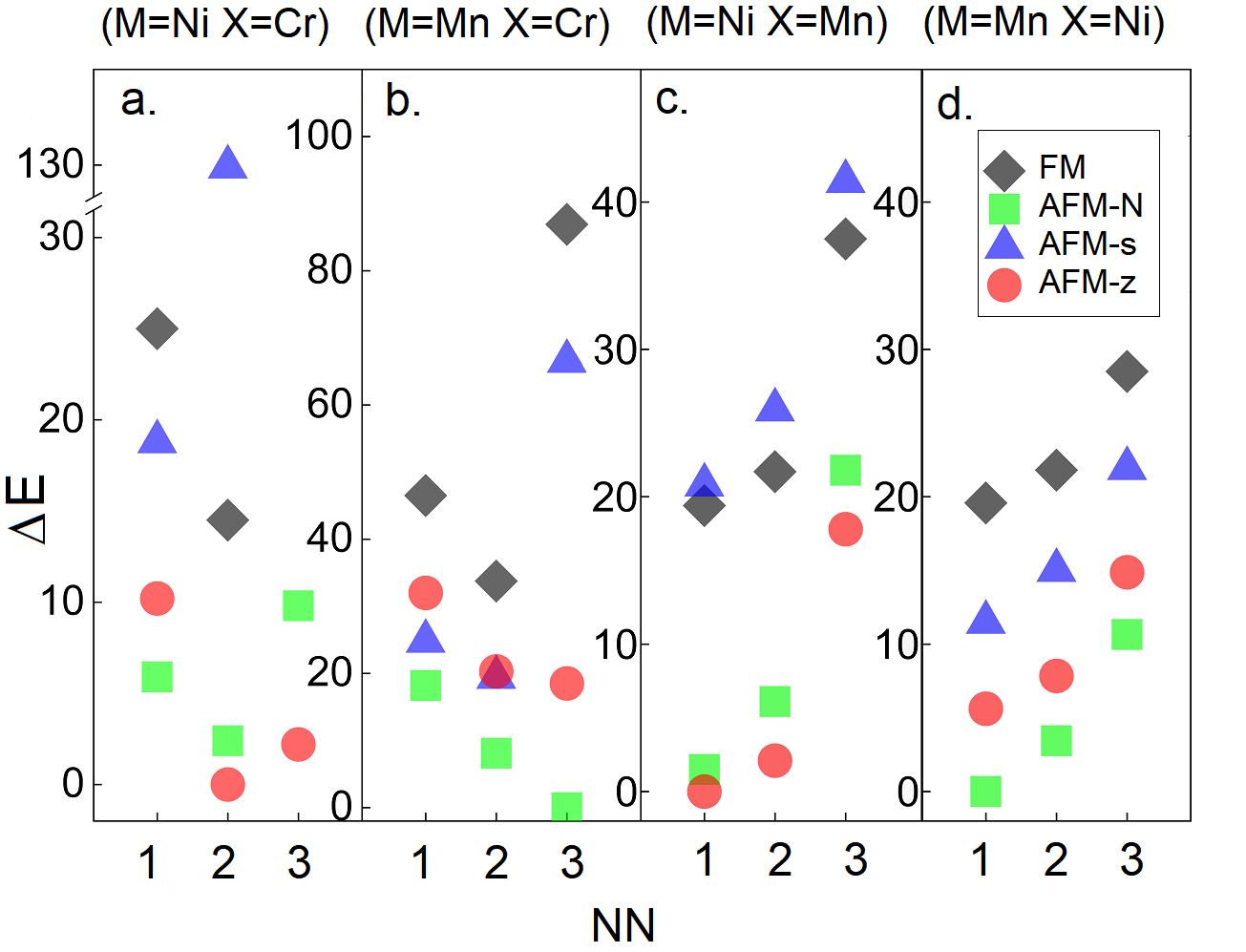}
\caption{\label{energy} a-d Energy difference ($\Delta E$) between a particular magnetic configuration (grey rhombus FM, green square AFM-N, blue triangle AFM-s and red circle AFM-z) and the magnetic ground state for each of the alloys (M$_{3/4}$,X$_{1/4}$)PS$_3$. Note, that for each of the plots, the most energetically preferable arrangement of the dopants is at the lowest energy. The energy is given in eV per magnetic atom for various structural arrangements of the dopant in the host (1NN, 2NN, 3NN).}
\end{figure}
\begin{figure*}
\centering
\includegraphics[width=1.0\textwidth]{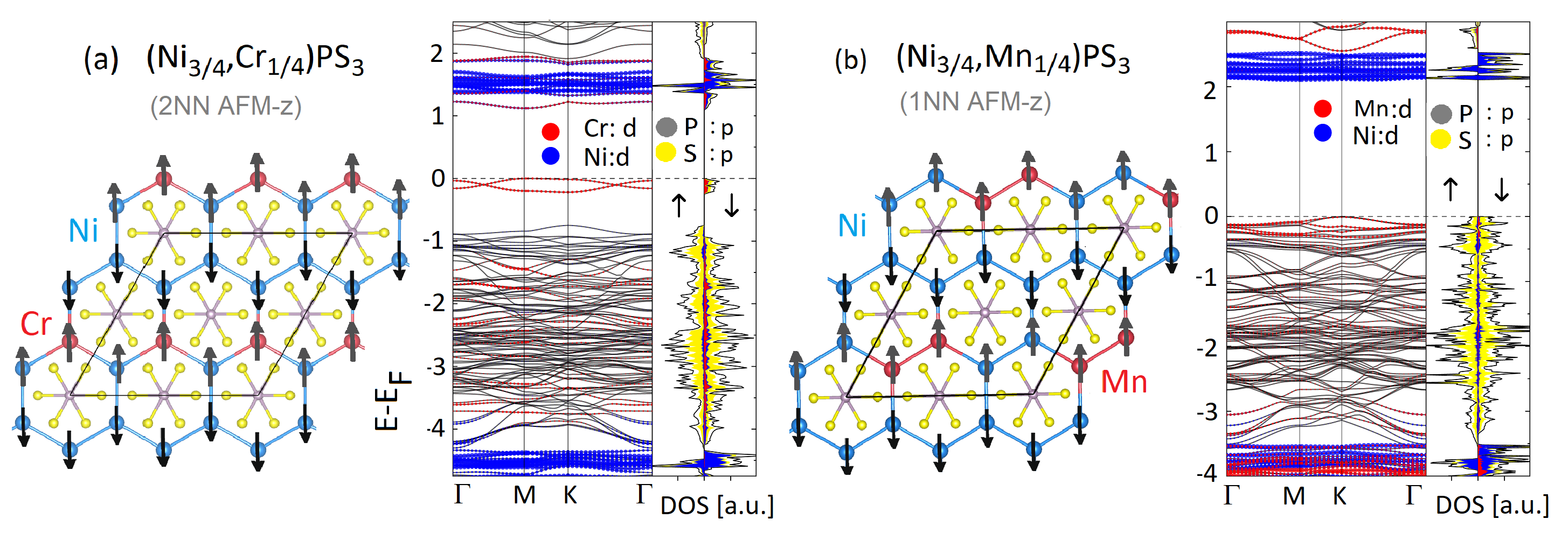}
\caption{\label{bands} Band structure and projected density of states (PDOS) for 2D ferrimagnetic alloys: (a) (Ni$_{3/4}$,Cr$_{1/4}$)PS$_3
$ and (b) (Ni$_{3/4}$,Mn$_{1/4}$)PS$_3$. On the left side of (a) and (b), the corresponding atomic structural arrangements and local spins magnetic configurations are presented. Note, that the spin of the dopant and the host atoms have different magnitudes, and are not visible here (spin arrows are not in scale).  For better visibility, the spin channels and projections of the  \textit{p} states of S and P atoms are presented only in PDOS. The energy is given in eV and Fermi level is placed at zero.}
\end{figure*}

\begin{table}[]\footnotesize
\caption{Results for the most energetically favorable position of the dopants in the (M$_{3/4}$,X$_{1/4}$) alloys. M is the total magnetization of the cell.  } \label{tab:electr}
 \def\arraystretch{1.5}
\begin{center}
\begin{tabular}{  c|  c|  c|  c}
   \hline
alloy & magn. state & M [$\mu_B$]  & band gap [eV] \\
 (M$_{3/4}$,X$_{1/4}$)PS$_3$  & $(2\times2)$ sc & per sc& \\
\hline
 (Mn$_{3/4}$,Cr$_{1/4}$)PS$_3$ &3NN AFM-N  & 0 & 1.7  (D, $\Gamma$)\\
 (Mn$_{3/4}$,Ni$_{1/4}$)PS$_3$ &1NN AFM-N & 0& 2.16 (D at K)\\
 (Ni$_{3/4}$,Cr$_{1/4}$)PS$_3$ &2NN AFM-z & 3.9 & 1.1 ($\uparrow$, D) 2.11 ($\downarrow$, ID)\\
 (Ni$_{3/4}$,Mn$_{1/4}$)PS$_3$ &1NN AFM-z& 5.7 & 2.11 (for both spins) \\
\hline
\end{tabular}
\end{center}
\end{table}

Let us first discuss the impact  of the dopant on the energetic and structural properties of the hosts. The energy difference between the  particular
configuration and the magnetic ground state for each 
considered alloy  is presented in Fig. \ref{energy}. Note that each of the employed alloys preserves the magnetic ground state of the host, independently of the structural arrangement of the impurity, and the type of the dopants (see Fig. \ref{energy}). In addition, the favourite Mn and Ni dopant position is  at the 1NN  (see Fig. \ref{energy}c and d), whereas the Cr dopants prefer to lay further apart, in particular, at 2NN (see Fig. \ref{energy}a) and 3NN (see Fig. \ref{energy}b) structural positions for NiPS$_3$ and MnPS$_3$ hosts, respectively. This reveals that the Mn, Ni dopants have tendency to cluster, while the Cr ions prefer to be spread over the host. From now on, we only discuss the magnetic ground state configurations of the alloys (the one which exhibits the lowest energy in Fig. \ref{bands} for particular alloy). The magnetization and the band gaps of the of these systems are collected in Table \ref{tab:electr}.  
All considered alloys exhibit  semiconducting behaviour (see Table \ref{tab:electr}). However, only (Ni$_{3/4}$,Cr$_{1/4}$)PS$_3$ and (Ni$_{3/4}$,Mn$_{1/4}$)PS$_3$ alloys have a nonzero net magnetization. Thus, we focus our discussion to these two ferrimagnetic systems. Owing to the different spins of the host and dopant atoms, as well as particular arrangement of the dopants in the NiPS$_3$ host, the ferrimagnetic state appears in these systems. The band structure and the orbital projections of these two  ferrimagnetic alloys are presented in Fig. \ref{bands}. Note that in the (Ni$_{3/4}$,Cr$_{1/4}$)PS$_3$ case, bands close to the Fermi level are mainly composed of  $3d$ states of Cr dopants, causing a sizeable reduction in the energy gap compared to pure NiPS$_3$ (2.3 eV for $U$=6 eV).  In case of Mn impurities, the valence band maximum is mainly composed of Mn $3d$  states, whereas, the conduction band minimum consists of very flat bands of Ni atoms (see Fig. \ref{bands} B).
\begin{table}[h]
\small
\caption{\label{tab:exchangeXM}The exchange coupling strengths $J_{i}$  calculated for various alloy systems, implied by the Ising model. Positive, negative $J_{i}$ indicate AFM and FM ordering, respectively. }  
 \def\arraystretch{1.3}
\begin{center}
\begin{tabular}{  c|  c|  c|c |c}
   \hline
\diaghead{\theadfont Diag ColumnmnHead II }%
{(M$_{0.75},$X$_{0.25})$PS$_3$}{  $J_i$ [meV]}
 &\thead{$J_1^{XM}$} & $J_2^{XM}$ & $J_2^{X}$&  $J_3^{XM}$ \\
 \hline
(Mn$_{3/4}$,Cr$_{1/4}$)PS$_3$& 2.5 &0.1 &-1.1& -0.4\\
(Mn$_{3/4}$,Ni$_{1/4}$)PS$_3$&-0.7 & -0.5&-0.6 &2.0 \\
(Ni$_{3/4}$,Mn$_{1/4}$)PS$_3$ &-1.0&0.3 & 0.15 &4.9\\
(Ni$_{3/4}$,Cr$_{1/4}$)PS$_3$&2.2 &0.4  &-0.3 &2.3\\
\hline
\end{tabular}
\end{center}
\end{table}

Next, we examine the mixed exchange couplings between the metal atom of the host and the impurity $J_i^{MX}$ within the classical Ising Hamiltonian on honeycomb lattice for the smallest possible cell containing 25$\%$ dopant concentration (for details see Appendix A and Fig. \ref{rectcell}). Note, that the FM exchange couplings are obtained  for the Mn and Ni nearest neighbors. The different values for $J_1^\mathrm{MnNi}$ (-1 meV) and $J_1^\mathrm{NiMn}$ (-0.7 meV) stems from the different lattice parameters of the hosts (see Table \ref{tab:exchangeXM}). Although, the mixed exchange coupling obtained for the  (Ni$_{3/4}$,Mn$_{1/4}$)PS$_3$ correctly reflected the 1NN AFM-z ground state configuration for this system, the difference between the 1NN AFM-N and 1NN AFM-z is just 1.5 meV per magnetic ion (see Fig. \ref{energy} C.). Thus, owing to magnetic frustration between the Mn and Ni atoms at the honeycomb sites, due to the competition between the N\'eel and zig-zag  configurations, similarly to results reported in \cite{PhysRevMaterials.5.064413}. Moreover, the critical temperature is directly related to the strength of the exchange couplings. Generally, the mixed exchange couplings are smaller than the corresponding metal atoms of the hosts.  Thus, the critical temperature of the mixed structure is expected to be  smaller than for the corresponding pure system, which is in line with recent experimental reports on the Ni$_{1-x}$Mn$_x$PS$_3$ alloy \cite{PhysRevMaterials.5.064413},  and in series of the mixed systems studies where the suppression of T$_N$  temperature with dopant substitution are reported \cite{MASUBUCHI2008668, Goossens_1998,PhysRevMaterials.4.034411}. In addition, one  should expect the further reduction of T$_N$ temperature of the employed mixed systems, due to the possible disorder of the dopant atoms  in the host,  as reported in \cite{V2000, PhysRevMaterials.4.034411} which is  not accounted in our studies.

\begin{figure}
\centering
\includegraphics[width=0.5\textwidth]{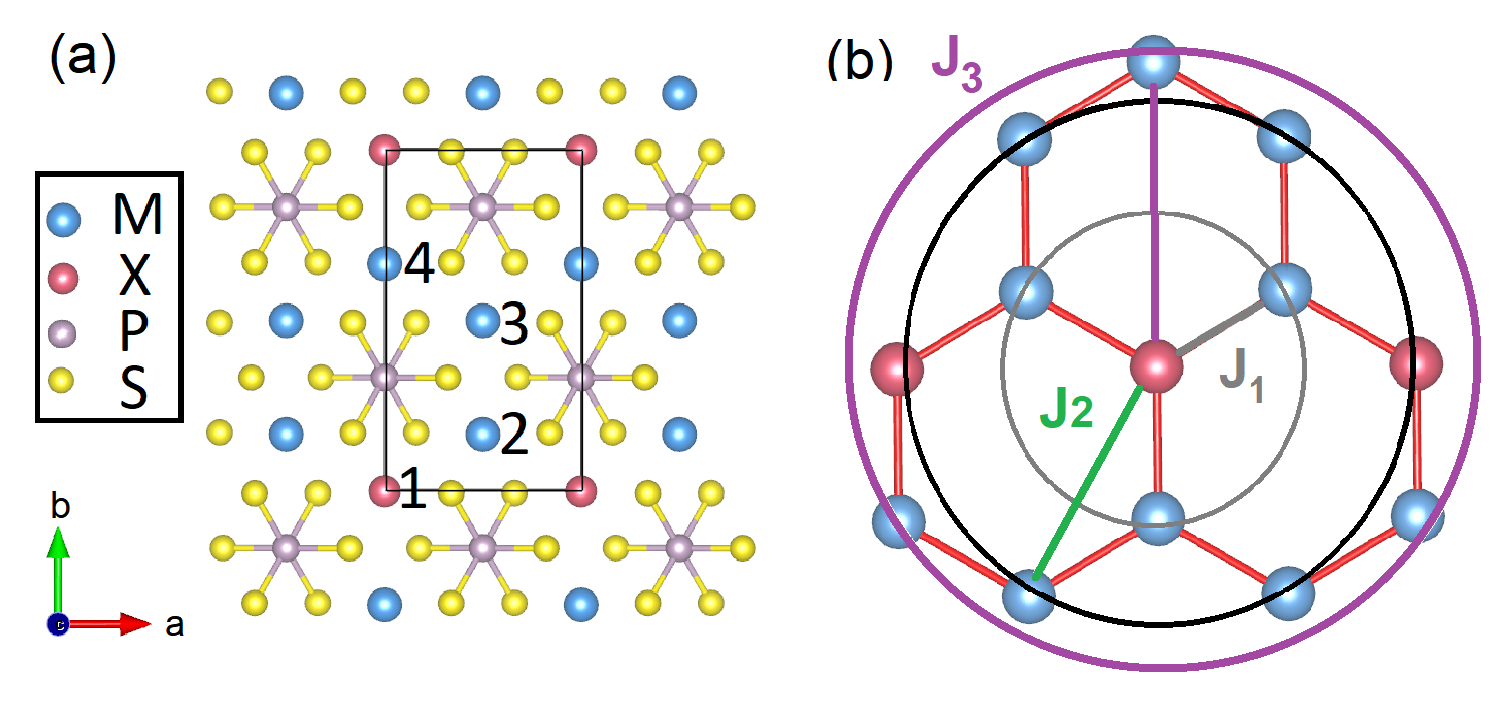}
\caption{\label{rectcell} (a) Rectangular planar cell denoted in black (smallest possible supercell for the impurity concentration of 25$\%$). (b) Schematic picture of the first (J$_1$, grey circle), second (J$_2$, green circle) and third (J$_3$, violet circle) nearest neighbour exchange couplings. Note that the spin arrangements for AFM-N, AFM-z, AFM-s are  1234=$\uparrow\downarrow\uparrow\downarrow$, $\uparrow\uparrow\downarrow\downarrow$, $\uparrow\downarrow\downarrow\uparrow$, respectively.}
\end{figure}

\section{Conclusions}
{\it First}, we examined the magnetic and electronic properties of the antiferromagnetic-ordered systems MnPS$_3$ and NiPS$_3$ (without magnetic impurities). 
We presented a qualitative explanation for the relative ratio of the different nearest neighbor (first-, second- and third-neighbor) exchange couplings by studying an effective direct exchange for both MnPS$_3$ and NiPS$_3$. In particular, we demonstrated that the third-neighbor exchange dominates in NiPS$_3$ due to the filling of the $t_{2g}$ subshell, whereas for MnPS$_3$ the first neighbor exchange is prevailed owing to the presence of the $t_{2g}$ magnetism.

We showed in this work that the onset of the nearest neighbor ferromagnetic coupling in NiPS$_3$ is due to the relatively strong (ferromagnetic) superexchange which is enabled by the complete filling of the $t_{2g}$ shell as well as by the (close to) 90 degree nickel-sulfur-nickel nearest neighbor bond. Nevertheless, even these relatively 
`fortunate' circumstances do not guarantee that the strong further neighbor exchange can also be ferromagnetic. 
The reason for that lies in the further neighbor bonds which are no longer of a `90 degree
variety'---and the latter bonds are, apart from specific Jahn-Teller effects, the only way to obtain ferromagnetic couplings in Mott insulators.
From a more general perspective, this study confirms the paradigm that the antiferromagnetic couplings are natural to the Mott insulating compounds.

{\it Second}, we examined the properties of the  MnPS$_3$ and NiPS$_3$ compounds doped with
25\% impurities (Ni in MnPS$_3$, Mn in NiPS$_3$, and Cr in both hosts). It turned out that all of the investigated alloys are Mott insulating, albeit with generally smaller band gaps than the corresponding host.
Crucially, we demonstrated an extreme robustness of the antiferromagnetic phases against impurity doping of the MnPS$_3$ and NiPS$_3$ compounds.
Hence, ferromagnetism cannot be easily stabilised by impurity doping, in agreement with the above paradigm. Nevertheless, as the dopants have different spins than the pure phases, the alloys exhibit ferrimagnetic properties for particular arrangements of dopants. 

The Mn and Ni impurities prefer to form dimers within the host, whereas the Cr dopants prefer to be further apart. Interestingly, unlike for the hosts, the first and second (dopant-host) exchange couplings are of similar order of magnitude. The latter leads to frustration in case of antiferromagnetic couplings.
We suggest that this may be one of the reasons 
of the observed lower magnetic ordering temperature of the doped systems \cite{PhysRevMaterials.5.064413}.

Our work sheds light on the origin of magnetism in the  antiferromagnetic family of  transition-metal  phosphorus trichalcogenides, by pointing out the  mechanisms which govern  the benchmark compounds, and  thus, extending  the fundamental  knowledge  of  2D magnetism.

\section*{Acknowledgements}
We acknowledge support by the  National Science Centre
(NCN Poland):
M. B. acknowledges support within grant UMO-2016/23/D/ST3/03446;
K. W. and C. E. A. acknowledges support within grant UMO-2016/22/E/ST3/00560;
K. W. acknowledges support within grant 2016/23/B/ST3/00839.
C. A. and G. C are supported by the Foundation for Polish Science through the International Research Agendas program co-financed by the
European Union within the Smart Growth Operational Programme. 
Access to computing facilities of PL-Grid Polish Infrastructure for Supporting Computational Science in the European Research Space and of the Interdisciplinary Center of Modeling (ICM), University of Warsaw is gratefully acknowledged, Grant No.~G73-23 and G75-10. 
We acknowledge the CINECA award under the ISCRA initiatives IsC76 ``MEPBI" and IsC81 ``DISTANCE" Grant, for the availability of high-performance computing
resources and support.
We made use of  computing  facilities  of  TU  Dresden  ZIH  within  the  project ``TransPheMat".

\begin{appendices}

\section{DFT Computational details}

The first-principle calculations are performed in the framework of spin-polarized density functional theory (DFT) as  is implemented  in  the VASP code \cite{PhysRevB.47.558,KRESSE199615}.
 The electron-ion interaction is modelled by using PAW psudopotentials \cite{PhysRevB.50.17953,PhysRevB.59.1758} with \textit{3s, 3p} states for P and S atoms, and \textit{3d, 4s} states for Mn, Ni, Cr treated as a valence states. The Perdew-Burke-Ernzerhof (PBE) exchange-correlation functional is employed \cite{PhysRevLett.77.3865}.
The kinetic energy cutoff for the plane-wave expansion of the  wave functions is set to 400 eV. A $\bold{k}$-mesh of $10\times6\times 2$ is taken to sample an irreducible first Brillouin zone of the rectangular planar cell (see Fig. \ref{rectcell}) containing 20 atoms including 4 transition metal atoms. The lattice parameters have been fully optimized within PBE+U approach for the magnetic ground state of the monolayers assuming the rectangular supercell. In particular,  the magnetic ground state (Hubbard U parameter) for MnPS$_3$ and NiPS$_3$ are AFM-N (U=5 eV) and AFM-z (U=6 eV), respectively. In the case of ($2\times2$) supercell which consists of 4 primitive hexagonal unit cells, the $5\times5\times1$ $\bold{k}$-mesh is chosen  to obtain the optimized position of the atoms. Considering, density of states calculations (DOS) the denser k-mesh equal to $10\times10\times2$ k-points is  taken into account. The convergence criteria for the energy and force are set to 10$^{-5}$ eV and 10$^{-3}$ eV/$\AA$, respectively. In order to properly model a monolayer system the 20 {\AA} of vacuum is added to neglect the spurious interaction between the image cells. Note that the standard  exchange-correlation functionals are insufficient to account for a non-local nature of dispersive forces, which are crucial for layered materials  and and adsorption molecules on
the surfaces\cite{PhysRevMaterials.2.034005,APP2011,AIPMilowska,commat2021109940}. Thus, the semi-empirical Grimme method \cite{Grimme} with a D3 parametrization is applied \cite{DFT-D3}.  For the 2D materials, we can use the HSE for a better description of the gap\cite{PhysRevB.104.Carmine}, however, in case of magnetism also GGA+U provides the same effect. We employ the GGA+U formalism  proposed by Dudarev \cite{PhysRevB.57.1505}  to properly account for on-site Coulomb repulsion between $3d$ electrons of transition metal ions, by using effective Hubbard U parameters.\\

Note that the proper choice of the U values is not straightforward due to the lack of accurate experimental information on the electronic properties. Also the common choice to compare the band gaps obtained in DFT with experiments to judge the U values also need caution, due to the fact that one-particle Kohn-Sham DOS cannot be directly compared to the measured data.
Thus, we decided to compute the  Hubbard U using  linear response method proposed by Cococcioni \cite{Cococcioni05} for the monolayers of MnPS$_3$ and NiPS$_3$, and we obtained 5.6 eV for the Ni and 5.3 eV for the Mn.
Therefore, we have used $U_{Cr}=4$ eV, $U_{Mn}=5$ eV and $U_{Ni}=6$ eV similar to the linear response results, which are typical $U$ values reported for MPX$_3$ materials in previous reports\cite{PhysRevB.94.184428}.
Moreover, our Coulombian repulsions are close to the typical values of $U$ in semiconductor compounds. Indeed, we find U=6.4 for Ni$^{2+}$ and $U=3.5$ eV for Cr in oxides\cite{PhysRevB.73.195107}, while the typical value of $U$ for Mn$^{+2}$ is $U=5$ eV\cite{autieri2020momentumresolved,Ivanov2016}.


\section{ Exchange interactions  calculated within DFT+U}

To examine the 2D magnetic structure of the employed alloys, we consider the classical spin Hamiltonian on honeycomb lattice, including  exchange interactions between nearest-neighbours (1NN), second nearest-neighbours (2NN), and third nearest-neighbours (3NN) between the metal atoms of the host (denoted as J$_i^{M}$ with i=1,2,3 as the number of neighbour) and between the host and dopant atoms (marked as J$_i^{XM}$ with i=1,2,3). 
In order to do it, we have chosen the smallest possible supercell (planar rectangular cell, see Fig. \ref{rectcell}) for the employed concentration. We insert the notation XMMM, in which the order indicates the first, the second, the third and the fourth atoms in the position represented in Fig. \ref{rectcell}. For the simplicity we skipped the letters in notation where applicable and we considered just the spins, e.g. $E_{\downarrow\uparrow\downarrow\downarrow}$ denoted the X atom with spin down, the M atom of the host on $2^{nd}$ $3^{rd}$ $4^{th}$ positions  with spin up, down, down components, respectively.

To extract the $J_i$ constants, we fix the lattice constants to that of the most energetically favorable spin arrangement of the host (AFM-N\'eel for MnPS$_3$ or AFM zig-zag for NiPS$_3$) and calculate the the energies for different spin configurations. We derive the following equations for all non-equivalent magnetic configurations:
 \begin{equation}
 \begin{footnotesize}
\begin{aligned}
E_{\uparrow\uparrow\uparrow\uparrow}=E_0&+(3J_1^{M}+7J_2^{M}+3J_3^{M}){|\vec{S_M}|}^2+ J_2^{X}{|\vec{S_X}|}^2\\
&+(3J_1^{XM}+4J_2^{XM}+3J_3^{XM}){|\vec{S_X}||\vec{S_M}|}\\
E_{\downarrow\uparrow\uparrow\uparrow}=E_0&+(3J_1^{M}+7J_2^{M}-3J_3^{M}){|\vec{S_M}|}^2+J_2^{X}{|\vec{S_X}|}^2\\
&+(-3J_1^{XM}-4J_2^{XM}+3J_3^{XM})|\vec{S_X}||\vec{S_M}|\\
E_{\downarrow\uparrow\downarrow\uparrow}=E_0&+(-3J_1^{M}+7J_2^{M}-3J_3^{M}){|\vec{S_M}|}^2+J_2^{X}{|\vec{S_X}|}^2\\
&+(-3J_1^{XM}+4J_2^{XM}-3J_3^{XM})|\vec{S_X}||\vec{S_M}|\\
E_{\downarrow\uparrow\downarrow\downarrow}=E_0&+(-J_1^{M}-J_2^{M}+3J_3^{M}){|\vec{S_M}|}^2+J_2^{X}{|\vec{S_X}|}^2\\
&+(J_1^{XM}+4J_2^{XM}-3J_3^{XM}){|\vec{S_X}|}{|\vec{S_M}|}\\
\end{aligned}\nonumber
 \end{footnotesize}
\end{equation}
\begin{equation}
 \begin{footnotesize}
\begin{aligned}
E_{\uparrow\uparrow\uparrow\downarrow}=E_0&+(J_1^{M}+J_2^{M}-3J_3^{M}){|\vec{S_M}|}^2+ J_2^{X}{|\vec{S_X}|}^2\\
&+(-J_1^{XM}+2J_2^{XM}+3J_3^{XM}){|\vec{S_X}|}{|\vec{S_M}|}\\
E_{\downarrow\uparrow\uparrow\downarrow}=E_0&+(-J_2^{M}-3J_3^{M}){|\vec{S_M}|}^2+ J_2^{X}{|\vec{S_X}|}^2\\
&+(-4J_2^{XM}-3J_3^{XM}){|\vec{S_X}|}{|\vec{S_M}|}\\
E_{\uparrow\uparrow\downarrow\uparrow}=E_0&+(-3J_1^{M}+7J_2^{M}-3J_3^{M}){|\vec{S_M}|}^2+J_2^{X}{|\vec{S_X}|}^2\\
&+(3J_1^{XM}-4J_2^{XM}+3J_3^{XM}){|\vec{S_X}|}{|\vec{S_M}|}\\
E_{\uparrow\uparrow\downarrow\downarrow}=E_0&+(-J_1^{M}-J_2^{M}+3J_3^{M}){|\vec{S_M}|}^2+J_2^{X}{|\vec{S_X}|}^2\\
&+(-J_1^{XM}-4J_2^{XM}+3J_3^{XM}){|\vec{S_X}|}{|\vec{S_M}|},\\
\end{aligned}
 \end{footnotesize}
\end{equation}
where $|\vec{S_X}|$ and $|\vec{S_M}|$ are the average spin magnetic moments of the atoms X and M, respectively. 
We assume the J$_i^{M}$ are calculated  from the host compounds (pure phases) and we compared with the similarly reported for the pure phases \cite{PhysRevB.91.235425}.  Then, for each of the alloy we have chosen its four lowest energies out of eight presented above, in order to extract the J$_i^{XM}$ and $J_2^{X}$ mixed exchange couplings.                              


\section{Wannier analysis}

To calculate the antiferromagnetic direct exchange for MnPS$_3$ and NiPS$_3$,
we extracted the real space tight-binding Hamiltonian with atom-centred Wannier functions with d-like orbital projections on the transition metals using the Wannier90 code\cite{mostofi2008wannier90,Pizzi20}. We calculated separately the hopping parameters for the orbitals symmetric and antisymmetric respect to the basal plane. The different symmetry and the separation in energy help to disentangle the two subsectors of the d-manifold\cite{Autieri2014NJP,Cuono_2019}. The calculation of the hopping parameters was done in the non-magnetic case to get rid of the magnetic effects and evaluate just the bare band structure hopping parameters, then the hopping parameters will be used for the model hamiltonian part\cite{PhysRevLett.127.127202}.
In order to have parameters to use for the model hamiltonian, we do not  perform the maximally localization so to have the Wannier function basis of our tight-biding model as close as possible to the atomic orbitals.\\

To calculate the competition between the magnetic couplings in NiPS$_3$ described in the formulas (\ref{direct})  and (\ref{SE}) of the main text, we have changed the strategy since we had to include the hybridization between the p-orbitals and antisymmetric d-orbitals (e$_g$ orbitals if we neglect the trigonal distortions).

\begin{figure}[h!]
\centering
\includegraphics[width=0.45\textwidth,angle=0]{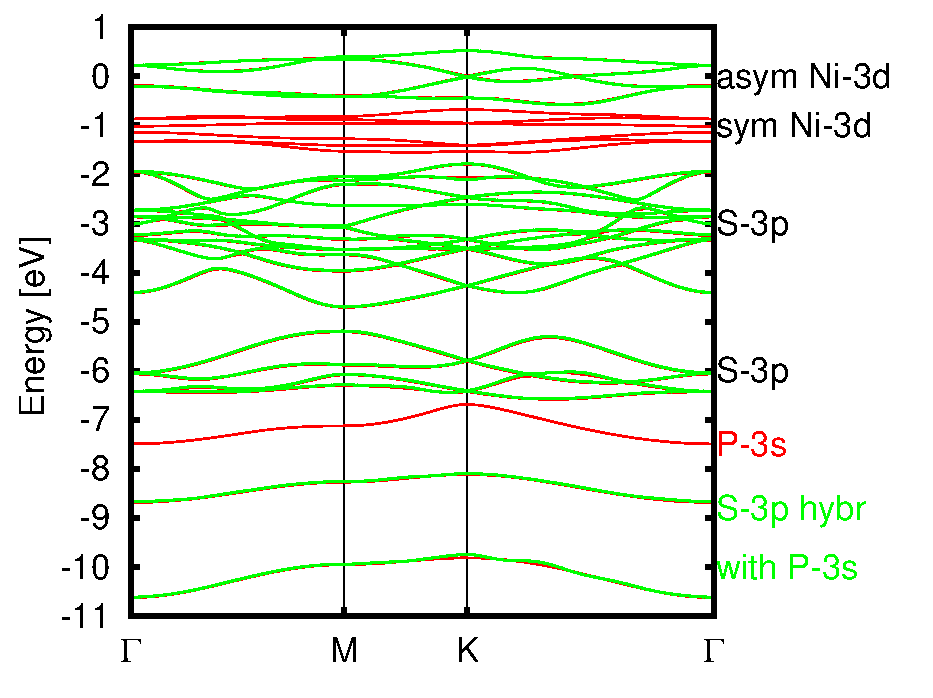}
\caption{\label{WannierNPS} Band structure of the nonmagnetic phase of the NiPS$_3$. Calculated GGA bands (red lines) and Wannier functions associated with the antisymmetric Ni-$3d$ and S-$3p$ orbitals (green lines). We labeled the bands with their main orbital contribution. The S-3p bands range from -2 to -11 eV with the two lowest bands that are strongly hybridized with the P-3s bands. The asym and sym Ni-3d states are the even and odd states, respectively.}
\end{figure}

In the simplified ionic picture of the NiPS$_3$,
we have Ni$^{+2}$, P$^{+4}$ and S$^{-2}$.
At the first-order of hybridization, we have the $pd$ hybridization between the $3d$-electrons of the Ni and $3p$-states of the S atoms and the bonding-antibonding between the $3s$-orbital of the P atoms. At the second-order of the hybridization, we have an hybridization between the 3s-orbitals of the P and 3p-orbitals of the S
forming the cluster PS$_3$ having oxidation state -2 and sp$^2$-orbitals. 
However, for our purposes, we can stop at the first order of hybridization, e.g., integrating out the $s$-orbital of the P atoms and  mapping the low energy band structure to a model Hamiltonian considering only the antisymmetric Ni-$3d$ and S-$3p$ orbitals.
We calculated the band structure of the nonmagnetic phase of the NiPS$_3$ and we plot it in Fig. \ref{WannierNPS}. The antisymmetric Ni-$3d$ bands are above the Fermi energy but they will be at the Fermi level once magnetism is turned on.  The bonding state of the P-$3s$ is at -7.5 eV from the Fermi level in Fig. \ref{WannierNPS} while the antibonding is above 2 eV (not shown). Among the occupied bands there are all the S-$3p$ bands and one P-$3s$ band.
The strength of the $sp^2$ hybridization is made clear from the similar shape of the bands between -11 and -7 eV in Fig. \ref{WannierNPS}.
We are able to map part of the band structure on a pd tight binding model, the electronic structure of the pd tight binding model is shown in green in Fig. \ref{WannierNPS} and it is in good agreement with the DFT band structure shown in red.
The numerical values of the relevant hopping parameters and difference in the energies on site $\Delta$ are reported in the main text.

\end{appendices}

\bibliography{main}




\end{document}